# TWO KINDS OF DISCOVERY: AN ONTOLOGICAL ACCOUNT

## Amihud Gilead

Department of Philosophy, University of Haifa, Eshkol Tower, Mount Carmel,
199 Aba-Khushy Avenue, Haifa 3498837, Israel
Email: agilead@research.haifa.ac.il

2 February 2014

*Abstract*. What can we discover? As the discussion in this paper is limited to ontological considerations, it does not deal with the discovery of new concepts. It raises the following question: What are the entities or existents that we can discover? There are two kinds of such entities: (1) actual entities and (2) possible entities, which are pure possibilities. The paper explains why the first kind of discovery depends primarily on the second kind. The paper illustrates the discoveries of individual pure possibilities by presenting examples such as the Higgs particle, Dirac's positron, and Pauli-Fermi's neutrino.

## 1. Some examples of two kinds of discoveries

The Standard Model, which is the model that physicists assume to describe correctly all the sub-atomic particles and forces working in nature, is strongly associated with some predictions and fascinating discoveries. Of particular interest are the sub-atomic particles W and Z, which are the weak force carriers, and the Higgs boson, the sub-atomic particle that is assumed to endow sub-atomic particles with mass. From now on, I will simply mention "particles" instead of "sub-atomic particles."

The 2004 Wolf Prize in physics was awarded to Robert Brout, François Englert, and Peter Higgs "for pioneering work that has led to the insight of mass generation … in the world of sub-atomic particles."[1] This has been considered as an

---

[1] See the site of the Wolf Foundation (2004). For clear and precise representations of the theory concerning the Higgs boson, consult Gross (2009) and Shears et al. (2006). On 8 October 2013, the Royal Swedish Academy of Sciences announced its decision to award the Nobel Prize in Physics for 2013 to François Englert and Peter W. Higgs "for the *theoretical discovery* of a mechanism that contributes to our understanding of the origin of mass of subatomic particles, and which recently was confirmed through

outstanding *discovery*, and the Wolf Foundation's announcement further states: "The discovery of Brout, Englert and Higgs was essential to the proof … that the theory with massive gauge particles is well defined; and subsequent calculations in that theory, verified experimentally, culminating in the discovery of the massive W and Z particles."

The particles called "gauge bosons" are the carriers of the fundamental forces of nature. Massive particles are those that have mass. Not all sub-particles have mass: photons, the light-particles, for instance, are massless. In contrast, all material particles must have mass. They gain their mass owing to an interaction with the Higgs boson. Without the Higgs particle, physicists are incapable of explaining why particles have mass. Thus, in the discovery of that particle hangs the fate of the whole of the Standard Model.

Two, quite different, kinds of discovery are mentioned in the Wolf Foundation's announcement as cited above: (1) the discovery of the Higgs particle and (2) the discovery of *actual* particles, W and Z, the carriers of the weak nuclear force, as they were verified experimentally and their masses were measured (in 1983 at CERN). Discovery (2) depended on discovery (1), for it is the Higgs particle that should endow W and Z with mass. We can easily understand what a discovery of an actual particle or entity is. Nevertheless, such was not the case of the Higgs particle until July 2012[2]: until then this boson was not discovered or detected as an actual

---

the discovery of the predicted fundamental particle, by the ATLAS and CMS experiments at CERN's Large Hadron Collider" (italics added).

[2] The CERN press release of 4 July 2012 announces: "… the ATLAS and CMS experiments presented their latest preliminary results in the search for the long sought Higgs particle. Both experiments observe a new particle in the mass region around 125-126 GeV." See the CERN site at http://press.web.cern.ch/press/PressReleases/Releases2012/PR17.12E.html.

particle (though in 2011 there were already some tentative empirical signs of its actual existence).

Before 1983, the status of W and Z particles was similar to that of the Higgs boson before July 2012—namely, they were not known then as actual entities, namely, as actualities. As expected, the CERN site defined particles, prior to their discovery as actual entities, as "predicted" or "hypothetical and novel" (CERN 1983). As this paper is an ontological account and for reasons that will be explained below, I suggest replacing "predicted particles" and "hypothetical and novel particles," which are quasi-epistemological terms, with "possible entities or existents," which are modal terms that fit well my ontological account. As I will suggest below, these possible entities are real as much as actual entities are, though differently.

Another illuminating example of the two kinds of discovery is that of the positron, the first discovered antiparticle. In 1928, Paul Dirac discovered it (namely, inferred its existence) *on purely theoretical grounds*, whereas Carl Anderson discovered the actual positron in experiments performed in 1932. Dirac referred to this particle as a positively charged electron, whereas Anderson named it "positron." Dirac's Nobel Lecture clearly shows how his notable equation opened up the *possibility* for the existence of a positively charged electron, which "one can infer" (Dirac 1933, p. 321) and which "appears not to correspond to anything known experimentally" (ibid., p. 323). Dirac characterized his discovery as an "inference" (ibid., p. 321) or "prediction" (ibid., p. 323).[3] Still, he was awarded the Noble prize

---

[3] Mark Steiner argues that this is a modern kind of prediction: "Prediction today, particularly in fundamental physics, refers to the assumption that a phenomenon which is mathematically possible exists in reality—or can be realized physically" (Steiner 2002, pp. 161–162). Such a non-deductive, "Pythagoreanized" kind of prediction is of the kind according to which "possible implies actual" (ibid., p. 162), and "[in] the case of Dirac's prediction, then, to predict the positron took courage or faith in mathematics. And the equation which supported this Pythagorean prediction

"for the *discovery* of new productive forms of atomic *theory*." On grounds of such theoretical considerations Dirac also predicted the following: "It is probable that negative protons *can* exist, since as far as the theory is yet definite, there is a complete and perfect symmetry between positive and negative electric charges, and if this symmetry is really fundamental in nature, it *must be possible* to reverse the charge on any kind of particle" (ibid., pp. 324–325; my italics. A. G.). It is a prediction of other antiparticles, whose *possibilities* the theory *necessitates* or infers.

In his Nobel Lecture in 1936, Carl Anderson, the discoverer of the actual positron, stated: "The present electron theory of Dirac provides a means of describing many of the phenomena governing the production and annihilation of positrons" (Anderson 1936, p. 368). In this case, too, the second kind of discovery, that of the actual positron, depends on the first kind, which is Dirac's discovery of the possibility of a positively charged electron. Before Dirac's discovery of this *possibility*, such an electron had to be considered as impossibility. It was not until Anderson's discovery of actual positrons that scientists changed their attitude toward this possibility and did not ignore or exclude it any more.[4]

The story of the discovery of the neutrino is even more fascinating. In a famous letter of 4 December 1930, Wolfgang Pauli reported that he had "hit upon a desperate remedy for rescuing" the compatibility of the law of the conservation of

---

… was 'derived' by purely formalist maneuvers" (ibid.). As the reader will see, the metaphysical view on which this paper is based sees this discovery differently—to begin with, the possible does not imply the actual; instead, the actual depends on the possible, which thus conditions the actual.

[4] For examples of the ignoring and misinterpreting of positron tracks before Anderson's discovery, see Segrè 1980 (2007), pp. 191–193. On the grounds of the dependence of the discovery of an actuality on that of its possibility, it is reasonable to assume that had those experimentalists paid enough attention to Dirac's discovery, which was about the possible existence of the positron, they could have, even before Anderson's discovery, correctly interpreted their findings instead of misidentifying them.

energy with the statistically empirical data concerning beta decay. The remedy was "the *possibility* [*Möglichkeit*] that there *might exist* … electrically neutral particles, which I wish to call neutrons" (Pauli 1994, p. 198; my italics, A. G.). In October 1933, Enrico Fermi reported on his hypothesis and theory concerning the beta decay and the existence of the neutrino (because two years after Pauli's discovery, James Chadwick discovered the neutron, Fermi suggested another name for the new possible particle—"neutrino," namely, "the little neutral one"). The experiments performed by Frederick Reines and Clyde Cowan in 1953 and 1955 detected an actual neutrino directly and, thus, their work "verifies the neutrino hypothesis suggested by Pauli and incorporated in a quantitative theory of beta decay by Fermi" (Cowan et al. 1956, p. 103). In this case, we encounter, to begin with, a discovery of a possibility, then a quantitative theory that establishes it by calculations and, finally, the discovery of the actual particle (which is not the end of the story, for there were some later discoveries concerning neutrinos, and at present the standard model comprises three kinds of them). Again, the whole story begins with the discovery of a *possibility*[5] on which the discovery of the actual particle depends. From now on, I will call it "Pauli-Fermi's neutrino possibility," although it was Pauli who referred to the neutrino as a possibility (according to Fermi, too); Fermi, on the other hand, referred to it as a hypothetical particle (or "the hypothesis of the existence of the neutrino"), awaiting an experimental confirmation.[6] Yet when mentioning Pauli's idea, Fermi refers to the

---

[5] Yet according to Reines, Pauli termed it also as a "postulate": "Pauli put his concern succinctly during a visit to Caltech when he remarked: 'I have done a terrible thing. I have postulated a particle that cannot be detected'" (Reines 1995, p. 204). I consider *Möglichkeit* in Pauli's letter as a preferable version, for it is a direct and authentic statement.

[6] Pauli expressed doubts as to the actual existence of the discovered particle. Hence he wrote in his letter of December 4, 1930: "I admit that my remedy may perhaps appear unlikely from the start, since one probably would long ago have seen the neutrons if

admitting of the existence of the neutrino as "a qualitative possibility" which squares facts concerning beta decay with the principle of the conservation of energy (Fermi 1933, p. 491). Unlike Dirac's discovery, Pauli's was of a *qualitative* possibility, namely it was independent of mathematical calculations and relied only upon theoretically physical considerations; whereas Fermi added the *quantitative* aspects to the discovered new *possibility*.

Bearing in mind these examples of discoveries, it is not clear at all, at least philosophically, *what* is the nature of the discoveries by theorists such as Dirac, Pauli, Fermi, Brout, Englert, and Higgs, which were *not* discoveries of actual entities or facts. Experimentalists discovered the relevant actual entities years later, whereas in the case of the Higgs boson until quite recently there was no decisive evidence of its actual existence. It emerges that the aforementioned theorists discovered some new possibilities. But what is the nature of these possibilities and what is the connection between them and actual reality? Are they merely possible entities? Or, were these not entities at all? And, if not entities or facts, *what* did these theorists really discover? Or, perhaps, were they not discoveries at all but inventions? Perhaps, then, these theorists simply invented, created, envisaged, or stipulated hypotheses, conjectures, or predictions? Or, after all, perhaps they discovered some entities, existents, or facts, unknown as actual at the time of those discoveries? If so, what kind of entities, existents, or facts did they really discover?

---

they existed" (Pauli 1994, p. 198). Having consulted Hans Geiger and Lise Meitner, he was more encouraged: "from the experimental point of view my new particles were quite possible" (ibid., p. 199). In contrast, the possibility of detecting such a particle was excluded by distinguished scientists such as Bethe and Peierls in 1934 (ibid.). Moreover, Niels Bohr pointed out in 1930 that no evidence "either empirical or theoretical … existed that supported the conservation of energy in this case. He [Bohr] was, in fact, willing to entertain the possibility that energy conservation must be abandoned in the nuclear realm" (ibid., p. 203). It is a typical way of abandoning or even excluding possibilities that would be discovered later as indispensable for scientific progress.

My account in this paper is mainly ontological. Thus, the discovery of concepts or ideas is not my present concern. Any discovery has to be of something, of some existent or entity. Indeed, concepts can be considered as discovered mental entities, but the discoveries I would like to discuss are of entities that are independent of our mind, whereas mental entities or concepts undoubtedly depend on our mind. Discoveries about ourselves, in the service of our self-knowledge and of knowledge in general—in philosophy, in psychoanalysis, and in other fields—are most valuable, but they are not my concern in this paper.

**2. Creation or invention**

Although the Higgs boson was not discovered as an actual particle until July 2012, as a possible entity it was neither a creation nor an invention. Brout, Englert, and Higgs's theoretical considerations and mathematical calculations implied the discovery of a new particle that the Standard Model had lacked. Thus, the Higgs boson has completed the description of the behavior of all sub-atomic particles and fundamental forces in nature. This comprehensive description should correspond to reality or nature existing independently of the theory. As taking a necessary part in such a description, the Higgs boson cannot be considered as an invention; it is a discovery. Neither Dirac's positively charged electron nor Pauli-Fermi's neutrino possibility were creations or inventions. They were, however, possibilities whose discovery led to the discoveries of actual particles. Similarly, scientists expected that the discovery of the Higgs boson as a possibility would lead to the discovery of the actual particle (namely, that this particle would be experimentally detected or empirically observed).

Creation or invention is quite different from discovery. Creation or invention produces its objects, which are entities that did not exist before, whereas discovery is of quite a different nature: it concerns what existed before, independently of the discoverer, as the discoverer does not create his or her discovery. In contrast, the products of creation or invention necessarily depend on the creator or inventor (in many of the cases, on the individual creator) and they could not exist without him or her, whereas the existence of the discovered entities or facts is independent of the discoverer in general and of any individual discoverer in particular. In the natural sciences and in mathematics, we can find some examples of several independent discoverers of the same discovery (the abovementioned example of Brout, Englert, and Higgs's discovery illustrates this perfectly).

**3. Conjectures, hypotheses, or predictions**

Conjectures, hypotheses, or predictions may be quite common and helpful in scientific discoveries. Major philosophers of science have devoted much thought to the contribution of conjectures and hypotheses to scientific knowledge (following Popper 1968). No less weighty appears to be the contribution of predictions to the discoveries of actual entities or facts.

Nevertheless, conjectures, hypotheses, and predictions are merely means to *discover* some entities or facts. The aim is the discovery, whereas the means to attain this end may be conjectures, hypotheses, or predictions. These are epistemological terms, whereas discovery is an ontological one. My question is: *What* kind of entities, existents, or facts did the abovementioned theorists discover? The answer should be in ontological terms, not in epistemological ones. Thus, their discoveries were not of a "hypothesized, conjectured, or predicted entities," which is a dubious hybrid of epistemological and ontological term (or a quasi-epistemological term). On these grounds, the Higgs boson, as a possible entity, was not a hypothesized, conjectured, or predicted entity; it was a discovered entity, participating indispensably in the Standard Model which describes mathematically the behavior of all sub-atomic particles and fundamental forces in nature. Similarly, by means of his equation, Dirac discovered a new particle—a positively charged electron. Such discoveries make predictions possible.

Although it is accepted to characterize the theory that makes it possible to understand and explain the origin of mass—first introduced by Peter Higgs (in 1963) and, independently, by Englert and Brout (in 1964)—as hypothetical in respect of the Higgs field as well as the Higgs boson,[7] I would like to attempt to characterize it differently. Both the Higgs field and the Higgs boson have been discovered entities or existents, though until quite recently they were not known as actual. As the Higgs field possibly permeates the whole universe, it is a discovered possible fact about the whole universe. Instead of "hypothesized facts," I prefer to use "possible facts." Equally, Pauli-Fermi's neutrino possibility should be considered as a possible entity. It was a discovered entity, not simply a hypothesis or conjecture, even though Fermi described it as hypothetical.

As I will argue below, possibilities can be legitimately considered as possible existents or facts for which ontological terms are valid and which are independent of the mind. Hence, possibilities are discoverable.

Because the aforementioned theorists had discovered possible entities, existents, or facts, they could predict the actual existence of such entities. Yet the prediction is not the discovery. It follows the discovery of the relevant possibility which precedes and conditions the discovery of the actuality in question. In any case, prediction is an epistemological term, not an ontological one.

---

[7] For instance, Shears et al. 2006, p. 3397.

9                                    Two Kinds of Discovery

**4. Fictions and thought-experiments**

Imagination plays a crucial role in constructing models which have contributed much to the making of discoveries. Model constructing may also employ fictions. Thought-experiments have been found quite useful for some major scientific discoveries,[8] and thought-experiments may consist of fictions. Such fictions, actually truthful fictions, serve us quite well in discovering real possibilities without which some of our most striking discoveries, if not all of them, could not be made.[9] The same holds true for the fictions involved in scientific models. These fictions are indispensable in serving scientists to make discoveries possible. We, thus, reach the same conclusion—imagination and fictions may serve scientists in achieving discoveries, both in theoretical and empirical or actual domains, but there is a major difference between these means and the discovered facts or entities, actual or possible.

**5. Conventions**

Is the discovery of possible entities simply a matter of convention? There were conventions about some alleged entities, for instance, phlogiston (in chemistry) and ether (in physics), and as soon as the conventions were discarded, no scientist believed any longer in the existence, possible or actual, of such entities.

---

[8] For instance, Szilard's discovery of the nuclear chain reaction, Rowland and Molina's discovery of the loss of the atmospheric ozone layer, and Mullis's discovery of the polymerase chain reaction (Hargittai 2011, pp. 244–245, 200, and 218–221).

[9] Regarding the vital role that truthful fictions play in discovering real possibilities, consult Gilead (2009).

Recently, Holger Lyre has raised doubt as to the reality of the Higgs mechanism as follows: "How is it then possible to instantiate a mechanism, let alone a dynamics of mass generation, in the breaking of … a kind of symmetry" which "is in fact a non-empirical or merely conventional one" and which does not possess any real instantiation, namely, realization in the world?[10] Entities that, to our knowledge, have no instantiations or realization, namely, actualization, in empirical reality are, in Lyre's view, mere conventions. Hence he assumes, wrongly, that non-empirical entities, which are possible entities, are, as a matter of fact, merely conventions. According to such a view, if the symmetries involved in the theory of the Higgs mechanism are about such entities, there is no real discovery involved, and it is simply a convention that so far physicists have accepted with no philosophical or otherwise critical grounds. This assumption is wrong, for the discovery of possible entities, which at the time of the discovery were not then known as actual (or for their actual existence there was no empirical evidence), can be quite real from a philosophical or scientific point of view, as I will argue, and it may be free of any convention or independent of it. Such is the lesson that I learn from the examples of Dirac's possible positron and Pauli-Fermi's neutrino possibility. Furthermore, in many cases, as in these two examples, such discoveries challenge the accepted views or conventions. Time will tell about the fate of the Higgs boson. Yet I see no reason why its discovery as a possible particle, a discovery made independently by different scientists, would be considered as a convention at all.

---

[10] Lyre 2008, p. 121. Lyre follows John Earman's skepticism that gauge or gauge symmetry is simply a "descriptive fluff," whereas philosophers of science should ask, "What is the objective … structure of the world corresponding to the gauge theory presented in the Higgs mechanism?" (Earman 2004, p. 1239). Likewise, Lyre emphasizes that the symmetry in discussion is "a merely conventional symmetry requirement" (Lyre 2008, p. 121). Thus, he reached the conclusion: "no ontological picture of the Higgs mechanism seems tenable, the possibility of an as-yet-undiscovered process or a mechanism … notwithstanding. But … the Higgs mechanism 'does not exist'" (ibid., p. 128). According to Lyre, the "possible existence of the as-yet-undetected Higgs boson … is a purely empirical question" (ibid., p. 130). The *possible* existence of the Higgs boson has *not* been an empirical question at all; however, its *actual* existence was such a question. It appears that Lyre assumes that actual or empirical facts are the only existing facts subject to discovery.

**6. Stipulation**

Discovery and stipulation exclude one another.[11] The existence of the Higgs boson is not a stipulation; it is a discovery, whether of a possible entity or of an actual one. If the latter, it was discovered by means of the powerful Large Hadron Collider (LHC) at CERN. Similarly, the discovery of the possibility of a positively charged electron was not a stipulation that Dirac's equation required; it was, however, a discovery of a real possibility which was inferred by means of a theory in general and an equation in particular. The same holds for Pauli-Fermi's neutrino possibility. It was undoubtedly discovered, not stipulated. When one stipulates, one does not mean to discover something or to put it to empirical test.

**7. Epistemic aids and the discovered existents**

Hypothesis, conjecture, prediction, fiction, thought-experiments, and the like all pertain to the epistemic aids for the discovery and should be discussed in the epistemology of discoveries. Yet the discoveries are not of these aids, the discoveries are of some *existents*, which are independent of these aids. It is impossible to discover something that does not exist, and it is meaningless to state that "one discovered something that does not exist," unless we would like to say that it was not a discovery at all but simply an illusion or fiction. "Existence" has at least two meanings, only one of which is actual.

---

[11] Hence, Saul Kripke claims: "'Possible worlds' are *stipulated*, not *discovered* by powerful telescopes" (Kripke 1980, p. 44). Cf. "Generally, things aren't 'found out' about counterfactual situation, they are stipulated" (ibid., p. 49). Contrary to Kripke, I think that individual pure possibilities are discovered, whereas fictions about them are possibly stipulated or invented.

Existents pertain to the ontic realm, which is philosophically investigated in light of ontological considerations. It is clear that the Higgs boson has been a purely physical existent of a special kind; until quite recently, it was not known as an actual existent. An actual existent is spatiotemporally and causally conditioned and it is empirically, directly or indirectly, observable or detectable. Until quite recently, there was no empirical evidence of the actual existence of this boson, though the Standard Model necessitates its existence—unless the Higgs boson existed, there was no explanation for the mass that each body or material entity must have. If no empirical evidence for the actual existence of this boson were found, this would have been pulled the ground from under the empirical validity of the Standard Model as a whole. Hence, there is an inseparable, necessary connection between the existence of the Higgs boson and the validity of the Standard Model as a whole. Similarly, there is a necessary connection between Dirac's equation and the existence of a positively charged electron, which is a discovered possible existent, and the positron, which is an actual particle; just as there is a necessary connection between Pauli's neutrino possibility, Fermi's theory concerning it, and the discovery of the actual neutrino.[12] The Higgs boson (as a possible particle), Dirac's positively charged electron, and Pauli-Fermi's neutrino possibility are discovered possible existents. What is the nature of such discovered possible existents? Are they similar to pure mathematical entities and the facts about them?

---

[12] In a similar vein, it is quite interesting to realize that Pauli wrote in a letter to Niels Bohr on February 15, 1955: "Einstein said to me last winter, … 'Observation cannot *create* an element of reality like a position, there must be something contained in the complete description of physical reality which corresponds to the *possibility* of observing a position, already before the observation has been actually made'" (Pauli 1994, p. 43; the italics are in the original).

There is a difference between pure mathematical entities and natural scientific possible entities such as the Higgs boson. Such possible entities or existents, unlike purely mathematical ones, are useless or insignificant in case that they have eventually no empirical or actual validity. Still, both kinds of existents, as we shall see, share something ontologically essential.

**8. Calculation and measurement**

Possible entities are not subject to measurement but to calculation, whereas actual entities—actualities— are subject to measurement, for, unlike possible entities, actualities are subject to empirical observation, spatiotemporal location, and causality. For instance, a point in Euclidean geometry cannot be measured, whereas a dot, an actual point, is measurable. In the Standard Model there are twenty-six parameters, describing the strength of forces, particle masses and so on, which "must be measured experimentally and then added to the model" (Shears et al. 2006, p. 3396). While the mass of some particles is very accurately predicted by *calculating* the binding energy of their constituents, until quite recently there was no corresponding theory which could predict the mass of the fundamental particles themselves and that of the Higgs boson itself (Shears et al. 2006, p. 3396). This has to wait for the experimental observations and *measurements* which are and will be performed at CERN. The masses of fundamental particles are ascribed to the existence of the Higgs boson. This actual existence is waiting for more empirical evidence which hopefully will also be achieved at CERN.

Calculations are *a priori* accessible and are primarily and directly valid for possible entities (such as the Higgs boson or a positively charged electron), to begin with, whereas measurements are only *a posteriori* accessible and are valid exclusively for actualities. Furthermore, calculations (such as Dirac's equation or Fermi's theory concerning beta decay and the neutrino) are associated with the necessity *about* the calculated facts (which are *possible* facts) and with the necessary relations among possible entities, whereas measurements have to do with actual entities and facts, which can be considered contingent (there is more about this below). Until quite recently, the predictions concerning the Higgs boson were based upon the correction of the calculations of the Standard Model, not upon measurements.

**9. Abstract or ideal entities**

Are possible entities or existents, such as geometrical entities and the facts about them, abstract or ideal entities? Although this is an accepted view about such entities, I consider it as wrong.

The received view is that geometricians in particular and mathematicians in general abstract from actual drawings of geometrical figures some ideal entities—"a circle," "a point," or "a line," for instance. By means of such abstractions they can make mathematical discoveries. In contrast, it is possible, following Kant or not, to show that mathematical discoveries are entirely independent of actual reality and empirical knowledge.[13]

---

[13] Giaquinto 2007, Ch. 4, "Geometrical Discoveries by Visualizing," shows how it is possible to make geometrical discoveries by visual means *in a non-empirical manner*. He thus relies on Kant in assuming synthetic *a priori* judgments in geometry (ibid., p. 50). Giaquinto's study is clearly epistemological, whereas I focus on the ontological aspects of discoveries. However, referring to Giaquinto's study, Daniel G. Campos discusses a similar view hold by Charles Peirce, for whom "reality is not circumscribed to what actually exists. 'Existing' and 'being possible' both are modes of 'being real'" (Campos 2009, p. 154). Campos relies at this point on Kerr-Lawson (1997). Kerr-Lawson, in turn, assumes that "no mathematical entities are existences in the fullest sense" (ibid., p. 79), as they are "hypothetical objects," and he somewhat connect this view with that of Putnam concerning "mathematics without foundations" (ibid., p. 84). In my view, in contrast, the entities that pure mathematics discovers,

Unlike its representation or image as an actual dot, a point, for instance, has a position but no dimensions (according to the first definition in Book I of Euclid's *Elements*); it cannot be measured and yet it exists "in" the Euclidean space, namely it is subject to an *a priori* order. A line (according to the second definition in that book) is a "length without breadth," whereas any actual drawn line must have some breadth, however small. Actual drawn circles, lines, or dots are subject to our observation, whereas pure circles, lines, and points are not; only their manifestations, depictions, phenomena, images, or representations in actual space are. Pure geometrical entities are thus "ideal," but this does not necessarily make them abstractions from actual reality. Neither are they idealizations of empirical facts, for if they were, they should be idealized according to some ideal standards or paradigms which, in turn, must be entirely independent of empirical facts and observation.

Actual drawn mathematical entities are actualities or depictions of purely possible mathematical entities. To identify actual mathematical entities we must rely upon such possible mathematical entities. Thus, to identify a dot as an actuality of a point we first must have access to the point as a possible entity. I argue this not on platonic grounds. I do not rely upon platonic paradigms or Ideas. I think about quite different entities, as I will explain below.

On such grounds, I do not consider mathematical entities as idealized abstractions from actual entities. Mathematical discoveries are, instead, of possible entities which are not idealized abstractions from actual ones but they precede anything actual.[14]

---

albeit purely possible, are real as much as actual existents are and they are not hypothetical objects. Both kinds of entities or existents are subject to discovery.

[14] Discussing mathematical discoveries, Gian-Carlo Rota relates to mathematical possibilities and proposes that "a rigorous version of the notion of possibility be

Mathematical proofs are necessarily valid for all possible relevant cases, whether actual or purely possible. Such cannot be the case of abstractions, however idealized, for abstractions first rely or, rather, are contingent, on some actual cases, from which they are abstracted, whereas the mathematical proof must be valid for *every* relevant possible case. Relying upon some actual cases, not upon all relevant possible cases, must make the case empirical and hence, contingent, and instead of a deductive proof we would rely only upon an inductive one. To assume that the inference and logics involved are *a priori* cognizable is not sufficient to substantiate the proof as universally valid, entirely independently of actual contingent cases, for logics is purely formal, whereas mathematics is different from formal logics, as it deals with contents and not with logical forms only. Hence, what makes pure mathematics exempt from actual constrains is not only its logical aspect; it is equally its purely mathematical aspect.

The discoveries of possible mathematical entities clearly show that they are primarily valid for possible facts that are associated with objective necessity, which cannot be ascribed to invention. That the sum of the angles of any Euclidean triangle is exactly $180^0$, for instance, concerns not only facts about any possible and actual Euclidean triangle; it also concerns the *necessity* about these facts. It is a necessary fact; there is nothing contingent about it. There is no Euclidean triangle that can be exempt from this fact. The mathematical discoverer must admit this necessity; he or

---

added to the formal baggage of mathematics" (Rota 1997, p. 191). In Rota's view, "[e]very theorem is a complex of hidden possibilities. … the proof of Fermat's last theorem foreshadows an enormous wealth of possibilities" (ibid., p. 195); or "proofs of theorems of Ramsey type are an example of a possibility that is made evident by an existence [non-constructive] proof, even though such a possibility cannot be turned into actuality" (ibid., p. 185). The happy expression "open up new possibilities for mathematics" (ibid., p. 190) in its various forms (ibid., pp. 191, 192, and 195) is a valuable leitmotiv in Rota's paper.

she is not entitled to free his or her mathematical way of thinking from it to invent any Euclidean triangle that is not subject to this necessity.

Purely mathematical entities are not purely physical entities. So what about our Higgs boson as a possible particle? Are my considerations about the existence of purely mathematical entities valid for possible ("purely theoretical") particles such as the Higgs boson? After all, if alas, no empirical evidence of the actual existence of this boson had been found, this possible entity would have become useless or insignificant for physicists. Such cannot be the fate of purely mathematical discoveries. Nevertheless, like purely mathematical entities, the Higgs boson was *not* an idealized abstraction from any empirical data or actual facts. Dirac's equation, Fermi's theory of the beta decay, the Big Bang model, and the Standard Model are not such abstractions. Instead, they comprise discoveries of possible entities and their relationality (the general term concerns all the ways in which entities relate one to the other). The existence of these entities has been independent of actual physical reality and it conditions the discoveries of the actual facts for which the models are valid. The possible existence of the Higgs boson is a necessary condition for its discovery and identification as an actual entity, which must be left to actual reality and empirical evidence. The possible existence of Dirac's positively charged electron was a necessary condition for its discovery and identification as an actual entity—the positron. It was Anderson who found the empirical evidence for the existence of the positron as an actual entity. The same holds for Pauli-Fermi's neutrino possibility and the discovery of actual neutrinos by Reines and Cowan. In each of these cases, the theorist's discovery of the possible particle opens the way for the experimentalist's discovery or detection of the relevant actual particle.

My view concerning the discovery of possible entities rests neither on platonic nor on Kantian grounds. The way I consider possible entities, mathematical or purely physical, is quite different, for it rests upon the idea of individual pure ("mere") possibilities and their necessary relations (in a general term—relationality).

## 10. Possible existents as individual pure possibilities

If possible existents are not idealized abstractions, then what are they? They are individual pure possibilities, which are real as much as actualities are, albeit differently. Regardless or independent of anything actual or of any actualization and exempt from any spatiotemporal and causal conditions, each individual possibility is pure. Individual pure possibilities are entirely independent of "possible worlds" as well as of any mind. The *concepts* of such possibilities are *de dicto*, but the possibilities themselves are possibilities *de re*.[15] As possibilities *de re*, individual pure possibilities are entirely independent of any mind and any concepts, and thus they are discoverable by us. We discover new individual pure possibilities, which are different from other pure possibilities, with some of which we are already familiar, and from known actualities as well.

To exist, any entity has first *to be* purely possible, to be a pure possibility. If an individual entity fails to exist as an individual pure possibility first, it cannot exist at all. Each existent, whether actual or not, has first to satisfy this *ontological* condition. Having this primary ontological condition satisfied and only then, the secondary ontological condition as to what are the spatiotemporal and causal circumstances under which this entity can or cannot actually exist, may or can be satisfied. Hence, the existence of any individual entity, whether actual or not, depends primarily on the existence of its pure possibility. Individual pure possibilities are thus the most fundamental existents.

---

[15] Thus, I do not follow Nicholas Rescher's conceptualism, replacing a "possibilism that is substantively oriented (*de re*)" by one that is "proportionately oriented (*de dicto*)." See Rescher (1999; 2003). For a critique of this powerful view see Gilead (2004). Nor I confine possibility to conceivability. There is much more to pure possibilities than conceivability, and the existence of individual pure possibilities does not depend on our mind.

No two pure possibilities can be identical—the law of the identity of the indiscernibles is necessarily valid for pure possibilities, which are exempt from spatiotemporal and causal conditions or restrictions. In other words, no two possibilities can be only numerically different, whereas two allegedly identical actualities can be only numerically different, for they exist at different places in the same time or at the same place in different times. Such cannot be the case of pure possibilities, which are exempt from any spatiotemporal restrictions; hence, the law of the identity of the indiscernibles is necessarily valid for them. Any "two" "identical" pure possibilities are, thus, one and the same possibility, and each pure possibility is an identity, too—the identity of the actuality that actualizes this pure possibility. Hence, below I will use the expression "pure possibilities-identities." With no access to the relevant pure possibilities-identities, scientists and laypersons alike may be doomed to be blind to the identity of phenomena or entities they may encounter.[16]

Because no two pure possibilities can be identical, each pure possibility is necessarily *different from* all the others. On this basis, each pure possibility necessarily *relates to* all the others. As a result, the realm of pure possibilities shares a universal unifying or systematic relationality.

---

[16] See, for instance, "We might marvel that Rutherford and Hahn did not grasp at the time the concept of isotopism, as they had discovered clear examples of isotopes; but when the mind is not prepared, the eye does not recognize" (Segrè 1980 [2007], p. 58). This is a good example of the indispensability of the discovery of a pure possibility-identity for the discovery and identification of the relevant actual entity or fact. At that time, Rutherford and Hahn did not consider the possibility that one and the same chemical element could have been two different physical entities. At the time, this was considered to be impossible.

In 1928, Dirac's discovery of the positively charge electron was of a pure possibility-identity, which was a necessary condition for the discovery of the actual positron. The same holds for Pauli-Fermi's neutrino pure possibility and the discovery of the actual neutrino.[17] Dirac's discovery of the pure possibility-identity of the positively charged electron and of those of other antiparticles predicted and paved the way to the discovery of the actual positron as well as other actual antiparticles. Until quite recently, the discovery of the Higgs boson was only the discovery of a pure possibility-identity. Without this fascinating discovery, physicists could not have predicted the actual existence of the Higgs boson, nor could they have explained how particles have mass, and how matter has been possible. As long as physicists had neither established evidence for the actual existence of the Higgs boson nor such evidence of its actual nonexistence, they still had well-established theoretical reasons to acknowledge its existence as a pure possibility. There appeared to be nothing to exclude it (despite some philosophical doubts). Thus, physicists thought that such a possibility *must exist* and should not be excluded; they knew *a priori* how to *identify* it; they understood and explained *why* it had to exist; they expected to discover its actual existence; and, thus, they *predicted* this discovery.[18] As a pure possibility-identity, the Higgs boson has been a necessary existent, owing to the relationality within the scope of the Standard Model.[19] This model necessitates the pure possibility-identity of the Higgs boson, and this necessity is independent of actual physical reality and empirical physical observations or experiments. In contrast, the physical utility and significance of this possibility depends, nevertheless, on actual physical reality and empirical observations or experiments, namely, on an empirical validity. The same holds true for the positron: the relationality of pure possibilities-identities involved in Dirac's theory and equation requires or necessitates the *possibility* of a positively charged electron regardless of actual reality and empirical observations or experiments.[20] Similarly, Fermi's theory and calculations about the beta decay and the neutrino *necessitates* Pauli-Fermi's neutrino possibility. Given that conservation of energy is valid for the beta decay, this particle *must* exist! Though "if you didn't see this particle in the predicted range then you have a very real problem" (Reines 1995, pp. 203–204). This reminds me very much of some quite recent thoughts about the Higgs boson.

---

[17] Frederick Reines entitled his Nobel Lecture (for the detection of the neutrino) in December 1995—"The Neutrino: From Poltergeist to Particle." In light of my view in this paper, I would like to rephrase this title thus: "The Neutrino: From Pure Possibility to Actual Particle." It is striking how both Pauli and Fermi were closely attached to Reines's detection of the neutrino. It was with Fermi whom Reines took advice since 1951 about the "possibility of the neutrino detection" (Reines 1995, p. 202; 206–208), and Pauli was the first to be informed by Reines's telegram about the detection: "We are happy to inform you that we have definitely detected neutrinos from fission fragments by observing inverse beta decay of protons" (ibid., p. 214). Pauli responded: "Everything comes to him who knows how to wait" (ibid.).

---

[18] As CERN Director General, Rolf Heuer, put it in July 2011, "We know everything about the Higgs boson except whether it exists."

[19] Although this model may have possible alternatives, the necessity under discussion holds true also, though differently, for the Higgs bosons in alternative models. Riccardo Barbieri, Lawrence J. Hall, and Vyacheslav S. Rychkov found it justified to consider possible alternative roads for physics beyond the Standard Model. See Riccardo Barbieri et al. (2006). For another possible alternative see T. Gregoire et al. (2004). Also consider Shears et al. 2006, pp. 3402–3403, for the Supersymmetry Model's prediction of five kinds of Higgs boson as well as for other possible alternatives.

[20] See Dirac 1928, p. 612. It was crucial that Dirac negated the attempt to exclude the very (pure) possibility of a positively charged electron. In this way, he opened up new possibilities for particles physics. On the way that excluding possibilities may result in blocking scientific progress and, in contrast, how saving possibilities contributes to this progress, see Gilead 1999.

Contrary to Steiner's interpretation (2002, p. 162), in Dirac's case possibility does not imply actual; instead, the actual is an actualization of pure possibilities, existing independently of our mind (and likewise in the case of the neutrino). By means of our theories, mathematical or physical, we gain access to these possibilities. The necessity in discussion is not necessarily a deductive relationality; there are many kinds of pure possibilities, each is necessarily different from the others, and each necessarily relates to the others. Dirac's discovery is about *physical* pure possibilities, which necessarily relate to one another and which are described *mathematically*. Mathematical description has been indispensable for any physical discovery since Galileo's days until our own. Whether Dirac was a "Pythagorean" (in Steiner's terms) or not, this does not necessarily reflect on his discovery of the positively charged electron, as long as we consider it as a pure possibility in the view that I present in this paper.

The relationality of mathematical pure possibilities in any mathematical proof is necessary, independent or regardless of any contingency and actuality or actualization. Unlike physically possible entities, the significance and strength of mathematical entities is independent of actual reality and empirical observations or experiments. Yet, physics, theoretical or applied, cannot exist without a strong reliance on mathematical language. Physics thus depends on mathematical pure possibilities and their necessary relationality. Pure mathematics and pure physical theory enable the discoveries of the pure possibilities-identities without which experimentalists cannot make the discoveries of the relevant actualities.

**11. How is *a priori* accessibility to pure possibilities possible?**

Our intellect and imagination are good enough to allow us access, however limited (as we are limited beings), to the realm of pure possibilities. Pure possibilities are certainly different from actualities. Observing actualities, we always can free our thought from actual constraints and think about—discover—pure possibilities, which are different from them. Our imagination and intellect help us to do so in many occasions. For some reason or other, we pay more attention to our capability of abstraction, and we are inclined to forget that we can abstract because we can think about pure possibilities that are different from the actualities with which we are already familiar. Moreover, we can think about pure possibilities that are quite different from the pure possibilities with which we are already familiar. On these grounds, we are capable of discovering new pure possibilities. In other words, we have accessibility to new pure possibilities even though and because they are different from all the actualities as well as of all the pure possibilities with which we are already familiar. We do not need empirical observations and experiments to have access to new pure possibilities. We can thus rely upon our thinking, intellect, and imagination to gain such access, which is certainly good enough to put forward our mathematical and pure scientific theories. Our accessibility to the realm of pure possibilities relies upon the universal relationality of all pure possibilities as well as actualities, insofar as they are actualization of the relevant pure possibilities-identities. Every possibility, pure or actual, including the possibility of one's mind, is different *from* the others and, hence, it relates *to* the others. This provides us with sufficient accessibility to the realm of pure possibilities, and this accessibility is *a priori*.

**12. Panenmentalism**

In the last fifteen years, I have introduced and elaborated on an original systematic metaphysics of a special kind, entitled panenmentalism or panenpossibilism (Gilead, 1999; 2003; 2009; and 2011). Until quite recently, I was not aware of the applications of this metaphysics to the discoveries discussed in this paper. Philosophy of science is one of the domains included in this systematically comprehensive metaphysics. This domain is the context in which this paper is embedded. Panenmentalism is a theory about individual pure possibilities and their universal relationality.

Panenmentalism is entirely different from any kind of possibilism known to me and it opposes actualism. The philosophical view that does not admit individual pure ("mere") possibilities altogether or at least as existing independently of actual reality—is called "actualism," whereas the view that does acknowledge such possibilities I term "possibilism." Actualism is generally allowed to use the idea of possible worlds and possible world semantics.

To the best of my knowledge, no actualist theory, including the most recent ones, admits the aforementioned absolutely independent existence of individual pure possibilities *or*, more traditionally, even any existence of them (consult, for instance, Bennett 2005 and 2006; Nelson and Zalta 2009; Contessa 2010; Menzel 2011; Woodward 2011; Vetter 2011, and Stalnaker 2004 and 2012).

Challenging actualism, panenmentalism is a possibilism *de re*, according to which pure possibilities are individual existents, existing independently of actual reality, any possible-worlds conception, and any mind (hence, they are not ideal beings). To the best of my knowledge, panenmentalism differs from any other kind of possibilism. Claiming that, it is not in my intention to argue that it is preferable to the other kinds; I say only that it is a novel alternative to them.

The following are the main features in which panenmentalism differs from other kinds of possibilism: First, Panenmentalism is strongly realist about individual pure possibilities, which are thus independent existents rather than mere "beings" or "subsistents." Over this point, panenmentalism disagrees with Meinonigians, Neo-Meinonigians (to begin with Richard Routely [Sylvan]; see Gilead 2009, pp. 23–27, 33–38, 46–47, 83–91, 109–113, and 121) and their followers (such as Graham Priest, Nicholas Griffin, Terence Parsons, and Edward Zalta). Second, it dispenses with the idea of possible worlds, which almost all the possibilists known to me have adopted. This idea is quite problematic for various reasons: for instance, it is not clear enough, and there are many controversies about it with no universal or long-standing consent; the problem of the epistemic accessibility from one world, especially from the actual world in which we live, to any other possible world does not appear to have a satisfactory solution; and if we can dispense with this idea and find a satisfactory, clearer and simpler, alternative to it, we should take this possibility into consideration. Third, panenmentalist pure possibilities are not abstract objects or entities, neither are they potentialities, for abstractions (as abstracted out of actualities or actual reality) and potentialities depend on actualities which are ontologically prior to them, whereas pure possibilities are ontologically prior to and entirely independent of actualities. Four, though using truthful fictions, panenmentalism, acknowledging the full, mind-independent reality of pure possibilities, differs from any kind of fictionalism, especially modal fictionalism (Gilead 2009, pp. 80–83; this difference holds also for Kendall Walton's make-believe theory). Five, as mind-independent, pure possibilities are not concepts, hence panenmentalism is possibilism *de re* and not conceptualism or possibilism *de dicto*.

If some readers may think that the panenmentalist pure possibilities allegedly remind them of Edward Zalta's "possible objects" or "blueprints" (Zalta 1983; and McMichael and Zalta 1980) or of Nino Cocchiarella's "possible objects" (Cocchiarella 2007, pp. 26–30; Freund and Cocchiarella 2008), such is not the case at all. First, these possible objects rely heavily on possible-worlds conceptions. Second, according to Cocchiarella's conceptual realism, framed within the context of a naturalistic epistemology, abstract intensional objects "have a mode of being dependent upon the evolution of culture and consciousness" (Cocchiarella 2007, p. 14), whereas panenmentalist pure possibilities are entirely independent of such evolution and of its naturalistic context as well and are *a priori* accessible. Third, following Meinonigians and Neo-Meinonigians, both Zalta and Cocchiarela consider "actual" and "exists" as synonyms, while, in their view, possible objects are merely "beings." In contrast, panenmentalism treats both pure possibilities and actualities as existents, though in different senses of the term "existence" (distinguishing between the existence of pure possibilities and that of actualities—the former is spatiotemporally and causally conditioned, while the latter is entirely exempt from these conditions).

Although David M. Armstrong adopts a special kind of possibilism (such as "possibilism in mathematics") and is committed to mere possibilities, namely, those without instantiation (Armstrong 2010, pp. 89–90), this is not a possibilist view in my terms: in Armstrong's view, these possibilities do not exist (ibid., p. 90), as he states that the only existence is spatiotemporal. Hence, his hypothesis is that there are no objects outside space-time (ibid., p. 5). Furthermore, though as a "one-worldler," Armstrong rejects other possible worlds (ibid., p. 16), the mere possibilities that he adopts explicitly supervene on the actual (ibid., p. 68). Thus, they are not pure possibilities in my terms (that is to say, entirely independent of anything actual). Finally, if mere possibilities are not existents, in what sense are they discoverable?

Since Yagisawa's modal realism heavily relies on the conception of merely possible worlds in which there are mere possibilia (Yagisawa 2010), I do not follow his view, either. The same holds for his distinction between "being" and "existence" or between "reality" and "existence." With panenmentalism, all individual pure possibilities are full-fledged existents—not only "real" ones. As for the problem of transworld identity, it does not exist for panenmentalism, avoiding the idea of possible worlds altogether.

With panenmentalism, pure possibilities are the possibilities-identities of actualities. Each actuality has a pure possibility-identity which cannot be shared with other actualities. This makes panenmentalism a unique kind of nominalism. Universal terms and laws rest upon the relationality of individual pure possibilities. Our accessibility to the realm of pure possibilities is *a priori*, whereas actualities are only *a posteriori* cognizable. Thus, our knowledge of actualities can be empirical only. Panenmentalism as a whole is thus neither empiricist, nor rationalist; yet it is rationalist about our knowledge of pure possibilities, and empiricist about our knowledge of actualities.

Necessity pertains to the existence of individual pure possibilities and to their relationality. Necessity also pertains to the inseparable connection between any pure possibility-identity and its actuality. There is no necessity at all about actualization. Thus, not all pure possibilities, albeit actualizable, are actualized, and the so-called "principle of plentitude" is *not* valid for actualities.[21] The contingency about each actuality is strongly compatible with the *a posteriori* and empiricist nature of our knowledge of actualities. The necessity about pure possibilities and their relationality can be *discovered* by means of logical, mathematical, and other theoretical considerations (including truthful fictions), but the discovered possibilities and their relationality are entirely independent of these considerations or means. As for our knowledge concerning the actualization of such discoveries, it is entirely subject to empirical observations and experiments.

---

[21] In contrast, Arthur Lovejoy's famous principle—"Possible implies actual" (Steiner 2002, p. 162) or "Any genuine possibility actualizes at some moment in an infinite time" (Bangu 2008, p. 249)— has been considered as inspiring the praxis of modern

But, if all actualities are contingent, what is the point in predicting actual existents on the grounds of pure possibilities-identities and their relationality? The crucial point is that only on the basis of such predictions can scientists empirically recognize, identify, understand, and explain the predicted actual entities. The discovery of the actual positron, of the actual omega minus particle, of actual particles such as W and Z, of some predicted actual elements in light of the eka-elements in the Periodic Table, and many other discoveries of actual entities are fine examples of demonstrating this crucial point. The *a priori* acquaintance with pure possibilities-identities made the discovery of the relevant actualities really possible. In contrast, excluding some possibilities on whatsoever grounds has hindered scientific progress (for instance, in the case of isotopes, the advent of quasicrystals [Gilead 2012] and others).

Pure possibilities are not ideal entities, which depend on our mind. We discover pure possibilities just as we discover actualities, though truthful fictions may help us greatly in discovering pure possibilities which are independent of our mind.

---

physics. For instance, Helge Kragh associates this principle, in its version as Gell-Mann's "totalitarian principle"—"Anything which is not prohibited [namely, possible] is compulsory"—with Dirac's reasoning (Kragh 1990, p. 272). On the aforementioned panenmentalist grounds, I see Dirac's discovery as well as the other discoveries discussed in this paper in quite a different light.

Thus, panenmentalism is not Kantian either. Pure possibilities and actualities are "things in themselves," not phenomena. This does not render our knowledge absolute or exempt from failure; on the contrary, although our accessibility to the realm of pure possibilities is *a priori*, our knowledge of it is quite limited. We know quite a little about pure possibilities and even more so about their *universal* relationality. It is inevitable that we are also subject to mistakes and errors about existents, possible or actual. After all, on grounds of "lazy" or convenient conventions, received views, preconceptions and so on, we quite habitually exclude vital pure possibilities, which are indispensable for our discoveries and scientific knowledge, and thus hinder scientific progress and fail in our aiming at truths. Nevertheless, because my discussion in this paper focuses on ontological considerations about discoveries, I do not discuss these major epistemological problems in this occasion.

Again, individual pure possibilities are not members of any possible world. Panenmentalism is exempt from possible-worlds semantic or metaphysics. As is well known, the idea of possible worlds has served actualists who have denied the existence of individual pure possibilities, which are entirely independent of actual reality in general and of actual individuals in particular.

**13. A Metaphysical Platform**

As all individual pure possibilities universally relate to each other, there is a metaphysical platform for embedding all there is in a universal system. On this platform, physical pure possibilities and their relationality also rest. The Standard Model reveals not only the symmetries that govern physical reality as a whole but also discovers how the breakings of these symmetries, which made it possible for particles to gain mass and to be material particles, are restored. Symmetry plays a crucial role in modern physics, not for aesthetic reasons and not necessarily for mathematical reasons,[22] I think, but because symmetry is a universal and unifying relationality of the multiplicity in nature under simple common law.[23] Note that the classical function of symmetry has been of "harmonizing" the *different* entities into a unified whole; whereas the modern concept relies *also* on equal entities, but still the relationality among the entities, whether different or equal, is maintained.[24] Panenmentalism bases the relationality of pure possibilities on their differences, as no two pure possibilities can be identical.

According the Standard Model, the massless photons can reach any point in the universe; they can spread themselves infinitely. Whenever the symmetry in the universe breaks, the omnipresent photons "mend" this and restore the symmetry. According to the panenmentalist metaphysical platform, photons thus *actualize* the basic universal relationality in the physical universe, for the photons communicate each distinct part of this universe to all the rest. The actualized relationality is the symmetry, apparent or hidden, that governs physical reality. We may say that the Standard Model thus discovers this symmetry in the two senses of discovery which this paper explicates: (1) the discovery of the relationality among all the physical individual pure possibilities (of particles and forces) and (2) the discovery of this symmetry as an actual fact about physical reality.

---

[22] Steiner argues that predictions by the use of symmetries "are of the (nonreductive) 'possible implies actual' variety because symmetry conditions define more what cannot occur rather than what must occur" (Steiner 2002, p. 162). According to panenmentalism, in contrast, because any actuality is contingent, no scientific prediction can be about "necessary" actual existence. The necessary relationality of pure possibilities does not imply actual necessity, whereas the inseparable connection between any pure possibility-identity and its actuality is necessary, though there is no necessity about the existence of any actuality. Given these restrictions, well-established predictions on grounds of symmetry may be very helpful scientifically, for instance, in the case of the discovery of the omega-minus particle. For an opposite view, questioning even the scientific status of such a prediction, which appears to be merely an educated guesswork, consult Sorin Bangu concerning the discovery of omega-minus (2008, pp. 256–257).

[23] For this reason symmetry has occupied the attention of physicists until the present: for instance, Pierre Curie's interest in crystals' symmetry; classical crystallography and the quasicrystals; symmetry's role in the special and general theory of relativity; in quantum mechanics; and in the Standard Model. One of the most illuminating insights of Pierre Curie was about the importance of symmetry in determining which phenomena are possible.

[24] Cf. Brading and Castellani (2008). As for permutation symmetry, it is a moot point whether the law of the identity of the indiscernibles is valid for quantum physics (ibid.), whereas panenmentalism applies it to every individual pure possibility.

31                                    Two Kinds of Discovery

**14. Two kinds of discovery**

In sum, there are two kinds of discovery: (1) discoveries of possible entities, which are individual pure possibilities-identities, and of their relationality; (2) discoveries of actualities. The second kind of discovery depends on the first kind. To discover new actualities we have to discover their pure possibilities-identities first or, at least, not to exclude these possibilities but to admit them, knowingly or unknowingly. One of the major hindrances in the path leading to scientific and other discoveries is our inclination to exclude possibilities from the outset. For instance, the discovery of quasicrystals was greatly hindered by the supposition that such crystalline structures were theoretically and empirically impossible.

    Were the positron and other antiparticles not purely possible in the light of purely physical theory, all that we know today in physics could be entirely different and some major discoveries of some actual antiparticles would not be possible from the outset. The same holds true for the Pauli-Fermi neutrino possibility and the application of the principle of the conservation of energy on the subatomic reality. Were the Higgs field and the Higgs boson not purely possible from the outset, namely, in the light of the purely physical theory of the Standard Model, physicists

could not attempt to discover their actualities, and our understanding, explaining, and knowledge of the universe would have been much less than they are today.